\documentclass[twocolumn,showpacs,preprintnumbers,prl,amsmath,amssymb,superscriptaddress, graphics]{revtex4-1}

\usepackage{amsmath}    
\usepackage{amssymb}
\usepackage[dvipdfm]{graphicx}   
\usepackage{array}
\usepackage{natbib}
\usepackage{url}

\newcommand{\SmB}{SmB$_6$ }

\newcommand{\BiSe}{Bi$_2$Se$_3$ }

\begin{document}

\title{Tuning bulk and surface conduction in topological Kondo insulator SmB$_6$}

\author{Paul Syers$^*$}
 \affiliation{Center for Nanophysics and Advanced Materials, Department of Physics, University of Maryland, College Park, MD 20742, USA}
 \thanks{These authors contributed equally to this work}
\author{Dohun Kim$^*$}
 \affiliation{Center for Nanophysics and Advanced Materials, Department of Physics, University of Maryland, College Park, MD 20742, USA}
 \affiliation{Department of Physics, University of Wisconsin, Madison, WI 53706, USA}
 \thanks{These authors contributed equally to this work}
\author{Michael S. Fuhrer}
 \affiliation{Center for Nanophysics and Advanced Materials, Department of Physics, University of Maryland, College Park, MD 20742, USA}
 \affiliation{School of Physics, Monash University, Victoria 3800, Australia}
\author{Johnpierre Paglione}
 \affiliation{Center for Nanophysics and Advanced Materials, Department of Physics, University of Maryland, College Park, MD 20742, USA}
 \affiliation{Canadian Institute for Advanced Research, Toronto, Canada M5G 1Z8}
 \email{paglione@umd.edu}

\date{\today}



\begin{abstract}
Bulk and surface state contributions to the electrical resistance of single-crystal samples of the topological Kondo insulator compound SmB$_6$ are investigated as a function of crystal thickness and surface charge density, the latter tuned by ionic liquid gating with electrodes patterned in a Corbino disk geometry on a single surface. 
By separately tuning bulk and surface conduction channels, we show conclusive evidence for 
a model with an insulating bulk and metallic surface states, with a crossover temperature that depends solely on the relative contributions of each conduction channel.
The surface conductance, on the order of 100~e$^2$/h and electron-like, exhibits a field-effect mobility of 133~cm$^2$/Vs and a large carrier density of $\sim 2 \times 10^{14}$~cm$^{-2}$, in good agreement with recent photoemission results. With the ability to gate-modulate surface conduction by more than 25\%, this approach provides promise for both fundamental and applied studies of gate-tuned devices structured on bulk crystal samples.
\end{abstract}

\maketitle


Recent theoretical work has proposed that the intermediate valence compound SmB$_6$ may be a member of a newly classified family of strong topological insulators \cite{Dzero104,Dzero85,Alexandrov111}. Called topological Kondo insulators, these systems differ from the conventional family of topological insulators \cite{Kane95,Fu98} such as Bi$_2$Se$_3$ because the bulk insulating band gap arises due to electronic correlations and opens at the Fermi energy. These materials are extremely interesting because of the potential for interplay between the topological states and other correlated electronic states, as well as the possibility to aleviate issues with chemical potential shifts due to intrinsic bulk doping \cite{Butch81,Qu329}.  

SmB$_6$, one of the first known Kondo insulator materials, has been of interest for many decades due to a long debate about the nature of its insulating state \cite{Allen20,Allen49}. 
It is now well known to harbor a $d$-$f$ hybridization gap that opens at low temperatures and has been well characterized by several experimental techniques to lie in the range of $\sim$10--20~meV 
\cite{Gorshunov59,Gabani52,Sluchanko61,Flachbart64,Zhang3,Yee1308,Roessler111,Ruan112,Xu88,Neupane4,Frantzeskakis3}.
The electrical resistance $R(T)$ of SmB$_6$ exhibits a thermally activated behavior at intermediate temperatures below room temperature, before saturating at an approximately temperature-independent value below a few degrees Kelvin \cite{Menth22,Nickerson3,Allen20,Caldwell75,Sluchanko61,Beille28,      Zhang3,Wolgast88,Kim3,Phelan4}. 
This robust property has recently been considered a key signature of topologically protected surface states \cite{Dzero104,Kim3}, prompting many experimental efforts designed to probe the topological nature of the conducting states in this material \cite{Li,Zhang3,Neupane4,Frantzeskakis3,Jiang3,Zhu111,Denlinger,Roessler111,Suga83,Nakajima1312,Neupane4,Xu88,Xu5}

Here we present resistance measurements probing the nature of surface conduction in bulk \SmB samples, using variations of bulk crystal geometry and surface ionic liquid gating techniques to, respectively, tune the bulk and surface conductance contributions. In both cases, $R(T)$ is well described by a low-temperature, temperature-independent surface contribution in parallel with a thermally activated bulk contribution, with a crossover temperature that depends on the relative values of each conductance component. 
Gate-tuned measurements using a Corbino contact geometry indicate a very large surface carrier density that can be dramatically changed by application of bias voltage. 
Our results strongly support the model of an insulating bulk with metallic surface states, as previously probed by other techniques \cite{Zhang3,Kim3,Wolgast88}, and characterize the tunability, mobility and carrier density of surface charge carriers, in good agreement with other spectroscopic techniques. 
Our study not only confirms the ability to tune the relative surface and bulk conductance contributions, but also paves the way for unique gate-controlled device construction on single-crystal samples of SmB$_6$.


Single crystals were grown using polycrystalline \SmB as the reactant and Al as the flux in a ratio of 1:200.  Starting materials were placed in an alumina crucible and sealed in a quartz ampoule under partial Ar pressure. Ampoules were heated to $1250^{\circ}$C and maintained at that temperature for 120 hours, then cooled at -2$^{\circ}$C/hr to 900$^{\circ}$C, followed by faster cooling.  Crystals were etched out of the flux using HCl, yielding mostly cubic-shaped crystals ranging in size from $\sim (0.2)^3$~mm$^3$ to $\sim (1.2)^3$~mm$^3$.  All samples were sanded and polished prior to contact placement.  Sample thickness was controlled by means of sanding and measured using an optical microscope, with uncertainties dominated by magnification resolution.  Thickness-dependent electrical resistivity measurements were performed using the standard AC technique, with four-wire geometry gold contact wires attached with silver conducting paint. Gating experiments were performed using a four-wire Corbino contact geometry pattern metalized with Au (200nm)/Ti (10nm) using thermal evaporation. After mounting on an insulating substrate, samples were covered with ionic liquid N,N-diethyl-N-methyl-N-(2-methoxyethyl)ammonium bis(trifluoromethylsulphonyl)imide (DEME-TFSI, Kanto Corporation) and an adjacent Au pad was used as a gate electrode. Ionic liquid application was performed inside a glove box and the sample was then transferred to the measurement cryostat within 5 minutes to minimize electrochemical reaction of DEME-TFSI with the ambient atmosphere. Gate voltage was applied in-situ inside the cryostat at $\sim 230$~K. Below the DEME-TFSI freezing point of $\sim 170$~K, the surface four-probe resistance was measured as a function of temperature. After each temperature sweep, gate voltage modulation was done by warming the sample to 230~K and changing gate voltage.


\begin{figure}
  \includegraphics[width=3.25in]{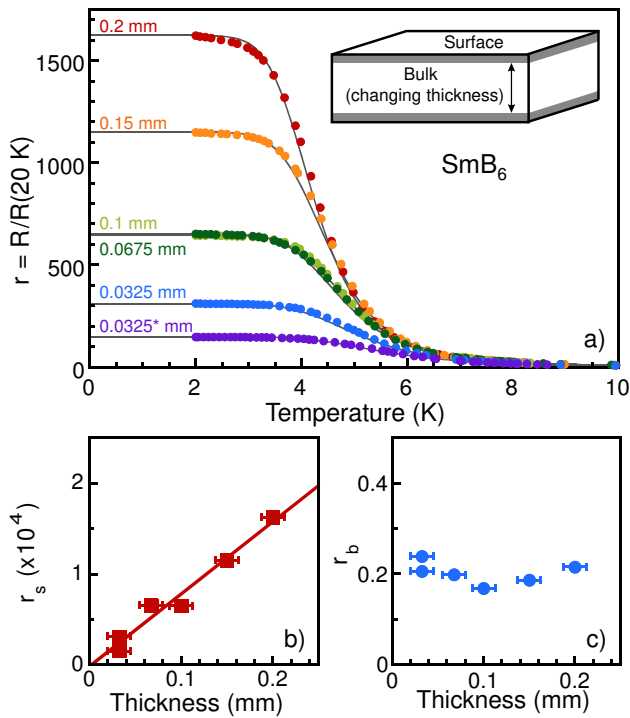}
  \caption{\label{Thickness} 
  a) Electrical resistance of a single crystal of SmB$_6$ as a function of sample thickness, normalized to its value at 20~K. The solid lines represent fits to the data using a two-channel conductance model [Eq. (1)], with fit parameters including the surface (panel b) and bulk (panel c) resistance ratio components $r_s\equiv R_s/R$(20~K) and $r_b\equiv R_b/R$(20~K), respectively, of the total conductance. [(*) Note geometry of sample for each 0.0325~mm thickness data set is slightly different due to loss of sample fraction.]}
\end{figure}

Fig.~\ref{Thickness}a) presents the temperature dependence of the longitudinal resistance of a single crystal as a function of thickness variation, with measurements taken subsequent to each thickness adjustment. To eliminate uncertainty in the geometric factor arising from varying contact geometry, we plot the resistance ratio $r\equiv R/R(20$~K), or $R(T)$ normalized to the resistance value at 20~K (approximately the temperature at which the Kondo gap is fully open). The resistance curves exhibit similar qualitative behavior to each other and to those reported in the literature \cite{Menth22,Kebede256,Gabani52,Wolgast88} over the entire temperature range up to 300~K, with 2~K resistivity values ranging between 0.5--2.9~$\Omega$cm. 

The crossover from high-temperature, thermally activated behavior to a low-temperature plateau in $R(T)$ has been interpreted as a transition from bulk state- dominated conduction to surface state-dominated conduction \cite{Kebede256,Kim3,Wolgast88}. This picture is consistent with the thickness dependence of resistance presented in Fig.~\ref{Thickness}a), which exhibits a clear separation of $r(T)$ curves from a single trace at higher temperatures to distinct plateau values for each thickness at low temperatures. In other words, the relative bulk-to-surface ratio of conductance shrinks with decreasing thickness, as expected due to the reduction of overall bulk conductance.

A simple parallel conductance model is used to extract the relative contributions, with total conductance described by $G = G_s + G_b$, where $G_s = 1/R_s$ is the surface contribution (assumed temperature-independent) and $G_b = 1/R_b$ is the bulk contribution, assumed to be activated in temperature due to a bulk energy gap $\Delta$.  
Therefore $G_b = W{\cdot}t/({\rho}_{b}L)e^{-\Delta/k_BT}$ with sample length $L$, width $W$, and thickness $t$; bulk resistivity ${\rho}_{b}$ in the high-temperature limit; and Boltzmann constant $k_B$. Thus, for the dimensionless and geometry-independent 
\footnote{We neglect a small dependence of surface conductance on changing surface area due to reduction in cross section perimeter.}
resistance ratio, 
\begin{equation}
r(T)^{-1} = r_s^{-1} + [r_b e^{-\Delta/k_BT}]^{-1},
\end{equation}
where $r_s \equiv R_s/R$(20~K) and $r_b \equiv R_b/R$(20~K) are the dimensionless, normalized surface and bulk resistance ratios, respectively. 

Fits to this model using $r_s$, $r_b$ and $\Delta$ as free parameters are shown as solid lines in Fig.~\ref{Thickness}a). For all thicknesses, we obtain a thickness-independent energy gap of $\Delta = 3.3 \pm 0.2$~meV, consistent with other transport measurements \cite{Menth22,Allen20,Sluchanko61,Beille28,Wolgast88}. 
The values of $r_s$ and $r_b$ are presented in Figs.~\ref{Thickness}b) and c), respectively, showing a clear contrast in their relationship with thickness;  $r_s$ exhibits a clear linear trend with thickness, while $r_b$ is independent of thickness. Understood in the context of their normalized nature, the linear relation of $r_s(t)$ translates to a linearly decreasing relative contribution of surface conductance compared to bulk conductance with increasing sample thickness. 
Conversely, the extrapolated value of $r_s(0)=0 \pm 0.00001$ at the zero-thickness limit translates to zero electrical conductance through the bulk, as expected in the bulk-surface model at low temperatures.

\begin{figure}[!]
  \includegraphics[width=3.25in]{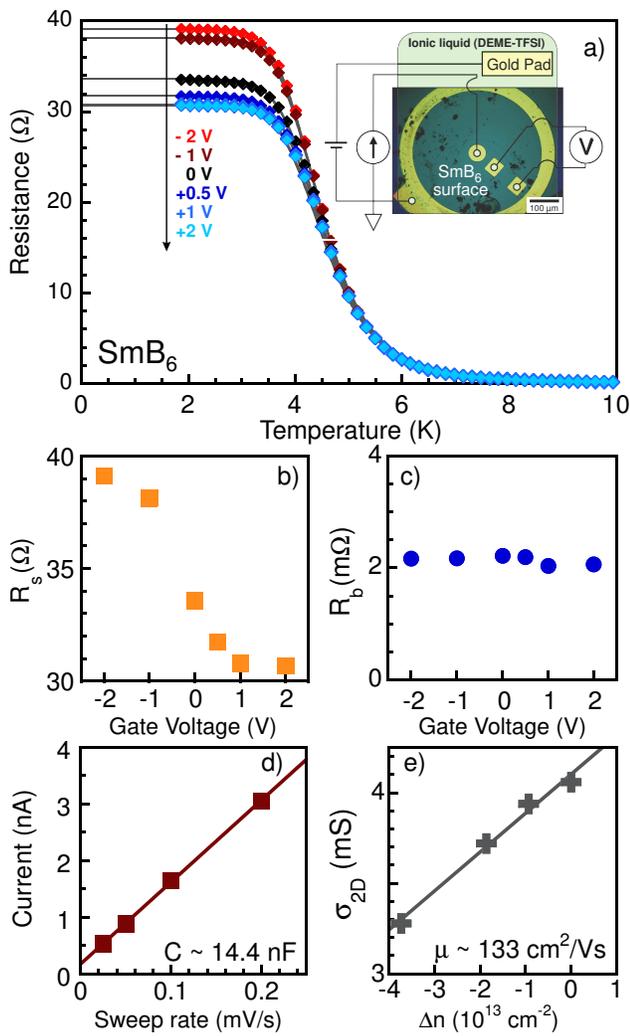}
  \caption{\label{Gating} 
  a) Electrical resistance of SmB$_6$ sample as a function of ionic liquid gate voltage, measured with a Corbino disk lead geometry placed on the surface of a single crystal, as shown in the inset diagram (superimposed on sample photo, with green shaded area representing the area covered by the ionic liquid gate structure).  Solid lines are fits to the two-channel conductance model [Eq. (2)].  
  Panels b) and c) present the surface and bulk resistance contributions, respectively, extracted from the two-channel fits as a function of gate voltage.
  Panel d) plots the transient current between the gate pad and ground as a function of gate voltage sweep rate, measured at 230~K. The solid line is a fit to a charging capacitor model with a capacitance of 14.4~nF. Panel e) shows the two-dimensional sheet conductivity as a function of the change in carrier concentration induced by gating. The solid line is a fit to a constant-mobility model with a mobility $\mu=133$~cm$^2$V$^{-1}$s$^{-1}$.    } 
\end{figure}

\begin{figure}[!]
  \includegraphics[width=3.25in]{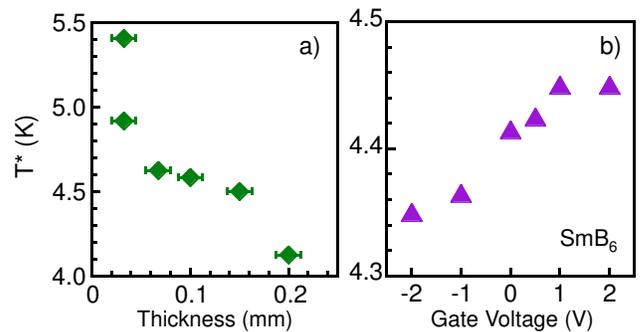}
  \caption{\label{Analysis} 
   a) Evolution of the crossover temperature $T^*$ between bulk- and surface-dominated conductance, defined as the inflection point in resistance temperature dependence as a fuctino of (a) crystal thickness variation and (b) ionic liquid gate voltage tuning. The non-constant evolution of $T^*$ is consistent with the two-channel conduction model (see text), showing the crossover depends solely on the relative bulk and surface contributions to the total electrical conduction.  }
\end{figure}


Previous experiments with electrochemical gating of \BiSe thin films have shown great success in shifting the Fermi energy from well within the bulk conducting band into the bulk gap, allowing the isolated Dirac surface states to be probed directly \cite{Cho11,Kim8,Kim4}.  Gating is, however, usually only effective at shifting the chemical potential of thin films or two dimensional systems, not bulk materials, as the gate electric field is confined to a thin region near the surface and heavily screened by bulk charge carriers. In the case of SmB$_6$, applying a gate voltage to surface of a bulk crystal is a simple, yet clear test of the surface vs bulk contribution of charge carriers. If the transport is dominated by surface conduction, the Corbino geometry ensures that the electronic transport occurs through one planar surface of the sample. 

For the IL gating measurements, a four probe Corbino geometry (see inset of Fig. 2) was patterned using e-beam lithography on a polished surface of SmB$_6$. 
Fig.~\ref{Gating}a) presents the $R(T)$ data for a single corbino device with various values of applied gate voltage $V_g$.  Similar to case of thickness variation (c.f. Fig.~\ref{Thickness}), variation of $V_g$ has no effect on $R(T)$ at higher temperatures, as exhibited by the collapse of all data onto a single trace above $\sim 5$~K. However, at lower temperatures a clear voltage-dependent splitting of $R(T)$ occurs, suggesting an identical tuning of bulk-to-surface contributions to the measured conductance, now controlled by a gate-controlled shift of the surface state chemical potential.

The same two-conductivity model can be applied, with the exception that resistance ratios are no longer needed since no geometries are varying. We therefore fit $R(T)$ to the following form:
\begin{equation}
R(T)^{-1} = {R_s}^{-1} + [R_{b}e^{-\Delta/k_BT}]^{-1}
\end{equation}
where $R_s$ is the (constant) surface resistance and $R_{b}$ the bulk resistance in the high temperature limit, and $\Delta$ is the gap energy as before.

Similar to the thickness case, we obtain a voltage-independent value of $\Delta= 3.78\pm 0.01$~meV. Presented in Figs.~\ref{Analysis}b) and c) are the results for $R_s$ and $R_b$, respectively, as a function of $V_g$, showing that the variation of $V_g$ has a dramatic effect on the surface resistance $R_s$, modulating it by over 25\% through the accessible voltage range, while the bulk resistance $R_{b}$ remains unaffected and relatively constant. Analogous to the thickness dependence of surface and bulk contributions, the observed relationships are strongly indicative of surface-dominated conductance in the low-temperature plateau region that can be modulated by variation of $V_g$. More important, this tuning directly confirms the surface-born origin of low-temperature charge carriers in \SmB and demonstrates the unique ability of controlling surface state conduction via device construction on the surface of a bulk cystal.

The Corbino gating experiment also provides information on the sign of charge carriers, their areal density and their mobility. As shown in Fig.~\ref{Gating}a), the variation of $R_s$ with $V_g$ is a decreasing function, consistent with the presence of dominant electron-like charge carriers at the surface. This may seem to be at odds with some measurements \cite{Sluchanko88,Kim3} and consistent with others \cite{Allen20,Cooley74,Wolgast88}, but it should be noted that traditional Hall effect experiments are no longer trivial in a situation with surface dominated transport, where conduction may be non-uniform. 
Recent angle-resolved photoemission measurements on \SmB have observed a surface state band structure consisting of both hole and electron pockets centered at the $\Gamma$ and $X$ points, respectively \cite{Xu88,Neupane4,Jiang3,Frantzeskakis3,Xu88}, suggesting a large difference in mobilities of each carrier type may be consistent with our observations. 
Using a simple capacitor model to fit the transient gate current dependence on $V_g$ sweep rate (Fig.~\ref{Analysis}d)), allows the determination of the gate capacitance (14~nF), or specific capacitance $c_g=3~\mu$F/cm$^2$, and hence the change in surface carrier density ${\Delta}n = c_gV_g/e$, where $e$ is the elementary charge. The measured sheet conductivity $\sigma_{2D}$ is approximately linear in the gate-induced change in ${\Delta}n$, as shown in Fig.~\ref{Analysis}e), indicating a constant field-effect and a surface carrier mobility of $133$~cm$^2$V$^{-1}$s$^{-1}$ (based on the measured resistance and the distance between voltage probe contacts -- see Fig.~\ref{Gating} inset). 
Extrapolation of the linear relationship to $\sigma_{2D}=0$ provides an estimate of  $n \approx 2 \times 10^{14}$ cm$^{-2}$ for the total carrier concentration. While the uncertainty of $\sim30\%$ for $\mu$ and $n$ is large due to difficulties in estimating sample area and geometric factor for the Corbino geometry, the absolute carrier density is in excellent agreement with recent photoemission results 
\footnote{Recent angle-resolved photoemission experiments generally observe a small ``$\alpha$'' band with Fermi surface area ~0.020(2$\pi/a)^2$ and two large ``$\beta$'' bands each with area 0.134(2$\pi/a)^2$ \cite{Xu88,Neupane4,Jiang3}. With a unit cell of (4.13 \AA)$^2$ and assuming each $\beta$ band is singly degenerate, this yields a total electron density of $\sim 1.7 \times 10^{14}$~cm$^{-2}$, which is very close to our estimate.}.
The relatively low surface state mobility then appears to be a natural consequence of the very low Fermi velocity for the surface states.

Finally, in both thickness and gating variation experiments above, the crossover from bulk- to surface-dominated conductance with decreasing temperature is observed to change as a function of the control parameter, as expected due to the change in relative weighting of each contribution. Fig.~\ref{Analysis}a) and b) present variation in this crossover temperature $T^*$, defined as the point of inflection of $R(T)$. The variation of $T^*$ with that of both bulk (thickness) and surface (gating) contributions to conductance confirms the two-channel model and disproves the prevailing idea that there is a static transition temperature; rather, it is merely defined by the relative contributions from the two conduction channels.

In conclusion, we have demonstrated methods to tune both the bulk and surface contributions to the electrical conduction, thereby providing conclusive evidence for surface-dominated transport in \SmB at low temperature. The evolution of transport with both sample thickness and surface gate tuning fits well to a two-channel conduction model involving a bulk, activated channel and a surface metallic channel. Furthermore, the measurement of gated surface conducting states using a Corbino lead geometry allows for the direct determination the electron-like sign of the surface charge carriers as well as the charge carrier density ($\sim 2 \times 10^{14}$ cm$^{-2}$) and mobility ($\sim 133$~cm$^2$V$^{-1}$s$^{-1}$ ). The charge carrier sign and density are in good agreement with previous photoemission results for \SmB. This study adds valuable information to our understanding of the proposed topological surface conduction in \SmB, and confirm the emerging picture that the surface of \SmB has a robust conducting state.

The authors would like to acknowledge R. L. Greene, X. Wang and Y. Nakajima for valuable discussion. Research at the University of Maryland was supported by AFOSR-MURI 
(FA9550-09-1-0603) and NSF (DMR-0952716 and DMR-1105224). M.S.F. is supported by an ARC Laureate Fellowship.

\bibliography{SmB6_Corbino}{}
\bibliographystyle{apsrev4-1}

\end{document}